\documentclass[journal]{IEEEtran}

\usepackage{graphicx}
\usepackage{subfig}
\usepackage{multirow}
\usepackage{array}
\DeclareGraphicsExtensions{.pdf,.png,.jpg,.eps}

\usepackage{cite}
\newcommand{\etal}{{\em et al.}}

\long\def\omitit#1{}
\begin{document}
\title{The Role of Trends in Evolving Networks}
\author
{\IEEEauthorblockN{Osnat Mokryn\thanks{Osnat Mokryn is with the School of Computer Science,Tel-Aviv Academic College, Tel-Aviv, Israel. Email: ossi@mta.ac.il} \hspace{2mm}
\and Marcel Blattner\thanks{Marcel Blattner is with the Laboratory for Web Science (FFHS), Zurich, Switzerland. Email: marcel.blattner@ffhs.ch} \hspace{2mm}
\and Yuval Shavitt\thanks{Yuval Shavitt is with the School of Electrical Engineering, Tel Aviv University, Israel. Email: shavitt@eng.tau.ac.il}}
	
}
\maketitle

%\section{Introduction}
\textbf{Modeling complex networks has been the focus of much research for over a decade \cite{watts1998small, albert2002statistical, newman2011structure}.
Preferential attachment (PA) \cite{barabasi1999emergence} is considered a common explanation to the self organization of evolving networks, suggesting that new nodes prefer to attach to more popular nodes. The PA model results in broad degree distributions, found in many networks, but cannot explain other common properties such as: The growth of nodes arriving late \cite{dorogovtsev2000structure} and Clustering (community structure). 
Here we show that when the tendency of networks to adhere to trends is incorporated into the PA model, it can produce networks with such properties. Namely, in trending networks, newly arriving nodes may become central at random, forming new clusters. In particular, we show that when the network is young it is more susceptible to trends, but even older networks may have trendy new nodes that become central in their structure.  Alternatively, networks can be seen as composed of two parts: static, governed by a power law degree distribution, and a dynamic part governed by trends, as we show on Wiki pages. Our results also show that the arrival of trending new nodes not only creates new clusters, but also has an effect on the relative importance and centrality of all other nodes in the network.
This can explain a variety of real world networks in economics, social and online networks, and cultural networks. Products popularity, formed by the network of people's opinions, exhibit these properties. Some lines of products are increasingly susceptible to trends and hence to shifts in popularity, while others are less trendy and hence more stable.  We believe that our findings have a big impact on our understanding of real networks.
}

Trends are a driving force in our lives. From economy to online engagement and popular research subjects, trends  govern many aspects \cite{denison1985trends, ng2003major, mathioudakis2010twittermonitor, asur2011trends}. Trends start locally \cite{Koenigstein:2012fk,dover2012network},  but only some of them, with time, create enough {\em buzz} to accumulate a global effect. We can think, for example, on a giant search engine that has started locally in the academic community at first. Alternatively, there is a popular recent Korean clip, that managed to become the number one hit on YouTube of all times\footnote{The references are to Google in the first example, and to the Korean song Gangnam Style in the second}.

We suggest that the preferential attachment process needs to be expanded to include the tendency of new and existing nodes to connect not only to rich nodes, but also to trending nodes. Hence, a node is {\em attractive} if it is either rich or trendy, and the relative weight of the trending factor is a characteristic of the network. We would like to accommodate our assumptions that networks have a tendency to trends, yet these trends may start randomly at any point. To this end, we extend the PA mechanism to include this tendency.

\textbf{Trending Preferential Attachment (TPA): }\\
We assume an evolving network where in each step a node is added with $m$ links.
Let a network's tendency to adhere to trends be denoted by $r$. Then, a node with degree $k$ that has acquired $\Delta k$ links in the last step will acquire new links in a
monotonically increasing rate that is a function of 

 \begin{equation}
 \label{tpa}
 k +  r \cdot \Delta k
 \end{equation}

The basic PA algorithm relies on random patterns. Once a node collects more incoming links it is more likely to become more popular. TPA builds on a similar idea: it looks at the {\em change} in the degree of a node, and amplifies this change by $r$, i.e., by the network's reaction to trends. The more trendy is the network, the bigger is the effect of small changes.
Hence, new nodes are more likely to attach to nodes that either have a high degree
or are gaining a momentum in the growth of the number of new links, and hence are {\em trendy}. By relying on random local changes and amplifying them, TPA is close in spirit to the PA model, unlike the static popularity model suggested in \cite{dorogovtsev2000structure}.

We next introduce a basic Trending Preferential Attachment growth algorithm.
Like in the PA model we start with $m_0$ nodes. Then, at each step, a new node  with $m\leq m_0$ links is added. The other ends of the links are chosen with a probability that correlates with the node's importance, denoted by its relative weight $W_i$:
\begin{equation}
\label{TPArule}
\Pi(W_i) = \frac{W_i}{\Sigma_j W_j}
\end{equation}
Where $W_i = k_i + r \cdot \Delta k_i$, and $\Delta k_i$ is the recent growth in the degree of node $i$ degree.  In the most simplistic way,   $\Delta k_i =\Delta k_i(t) = k_i(t) - k_i(t-1)$. 
At each step a node with $m$ links is added to the network, hence, the total weight at time $t$ is $2mt + mr$. The rate at which a node acquires edges is $\partial W_i/\partial t = W_i/(2t+r)$, which gives
\begin{equation}
W_i(t) = m\cdot (\frac{r+2t}{t_i})^{0.5}
\end{equation}
Where $t_i$ is the time in which node $i$ was added.

According to the TPA model, the $r$ parameter governs the trendiness of the network. For large $r$, some nodes may incur a sudden speedy growth, while for other nodes the fast growth stops. However, in slow changing networks, where $r$ is very small, the growth depends indeed much more on the degree of the node rather than on current trends.
Clearly, when $r \rightarrow 0$ then $W_i(t) = k_i(t)$.

In the thermo-dynamic limit, namely when $t \rightarrow \infty$, TPA differs from the PA model only by an additive factor: 
\begin{equation}
P(W_i<W) = P(t_i>\frac{m^2}{W^2}(r+2t))=1-\frac{m^2(r+2t)}{W^2(m_0+t)}
\end{equation}
Thus, $P(W)$ is given by
\begin{equation}
P(W) = \frac{\partial P(W_i(t) < W)}{\partial W} = \frac{2m^2(r+2t)}{(m_0+t)W^3} \rightarrow_{t\rightarrow \infty} \frac{4m^2}{W^3}
\end{equation}
Thus, we get $\gamma = -3$, as $W$ is equivalent to $k$ in the PA model.

\begin{figure*}[h]
\centering
\hspace{-.55in}\includegraphics[width=202mm]{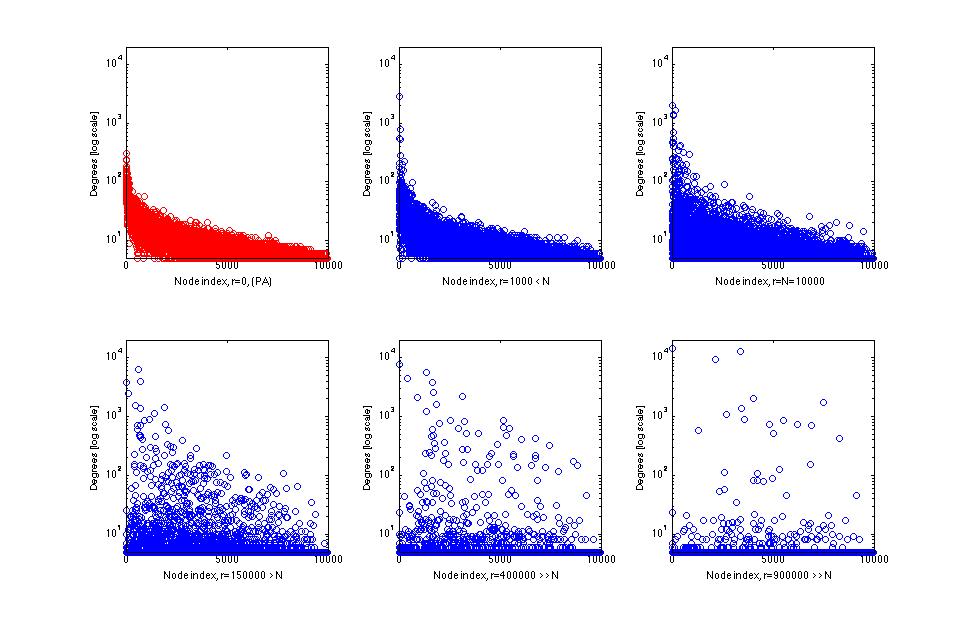}
\caption{Node degree as a function of the network trendiness $r$}
\label{fig:deg-r}
\end{figure*}

What we are interested in is the dynamics of young trending networks, namely, when $r >> N$, where $N$ is the number of nodes in the network. When the growth model allows for the addition of one node at each step, $N \sim t$.  Consequently, for very  trending young networks, we get :
\begin{equation}
\Pi(t) = \frac{W_i(t)}{\Sigma_j W} \rightarrow r^{-0.5}
\end{equation}
Showing that newly arriving nodes have a similar probability of becoming important as older nodes in the network.
Figure~\ref{fig:deg-r} shows the degree of nodes as the function of their arrival time. In the PA model, there is a strong correlation between the arrival time and the degree of a node, yielding the rich gets richer phenomenon. However, as the network's trendiness $r$ grows, the correlation weakens. For the case of $r=900000 >> N$, there is no correlation and high degree nodes can arrive almost at every time.

\begin{figure*}[h]
\centering
\begin{tabular}{ccc}
\hspace{-.3in}\subfloat[$N=2\cdot10^3, r=0$]{\includegraphics[width=6.8cm]{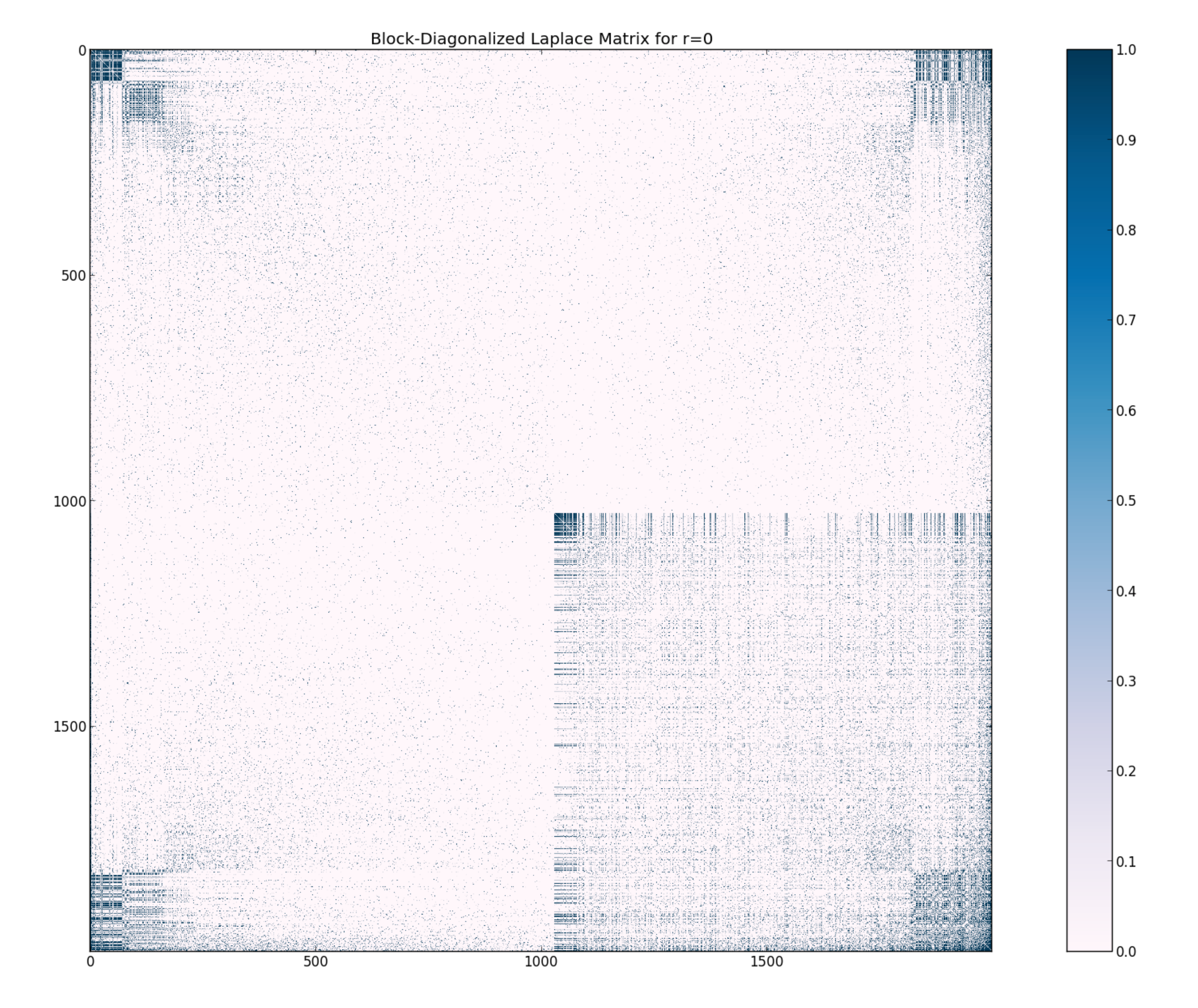}} &
\hspace{-.3in}\subfloat[$N=2\cdot10^3, r=2000$]{\includegraphics[width=6.8cm]{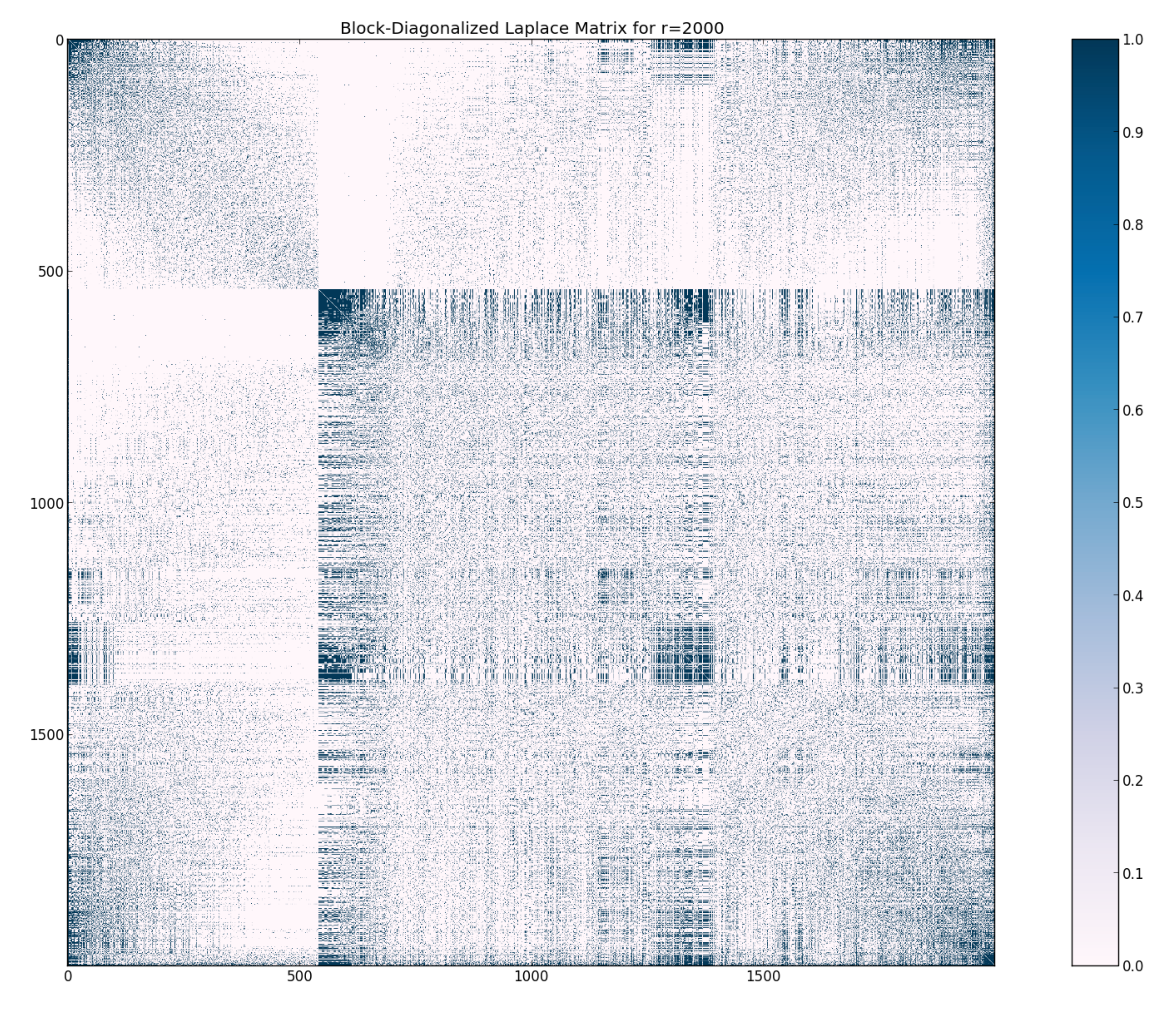}} &
\hspace{-.3in}\subfloat[$N=2\cdot10^3, r=2\cdot10^4$]{\includegraphics[width=6.8cm]{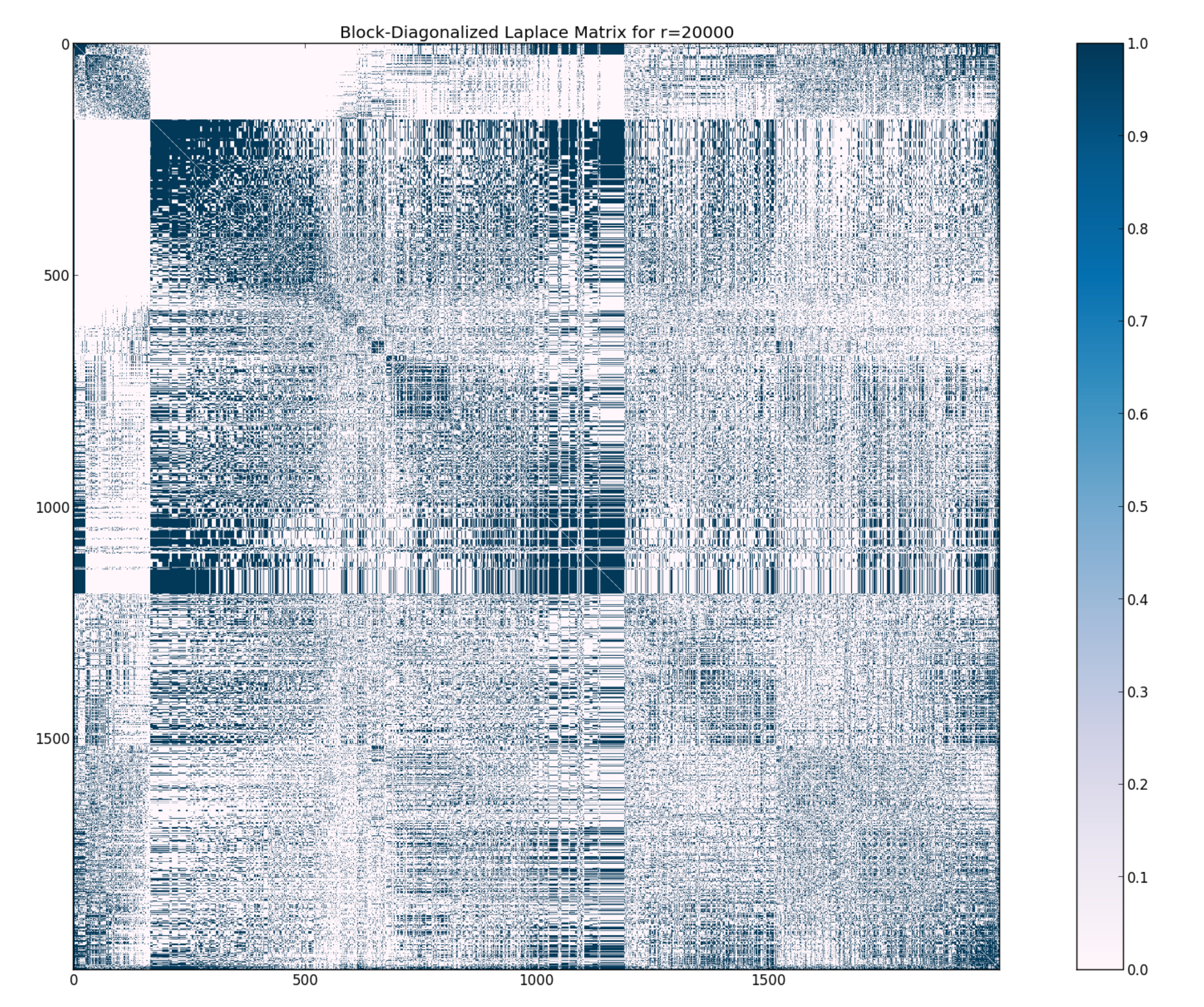}}\\
\end{tabular}
\caption{A heat map of the  block diagonalized Laplacian matrix. The graph produced by the corresponding Laplacian matrix L = D - A, where D is the degree diagonal matrix and A is the adjacency matrix.
In the diagonalized Laplacian matrix the eigenvalues are ordered in increasing order and hence also the eigenvectors (columns) accordingly. }
\label{fig:cluster-r}
\end{figure*}

The TPA model also reveals clustering behavior found in many real networks, unlike the PA model. Figure~\ref{fig:cluster-r} reveals the clustering behavior of networks as a function of $r$. Trending networks reveal high complexity relative to networks with a lower tendency to adhere to trends. Particularly,  with increasing r the structure becomes more complex and the number of clusters increases as well.

\begin{figure}[h]
\includegraphics[width=16cm]{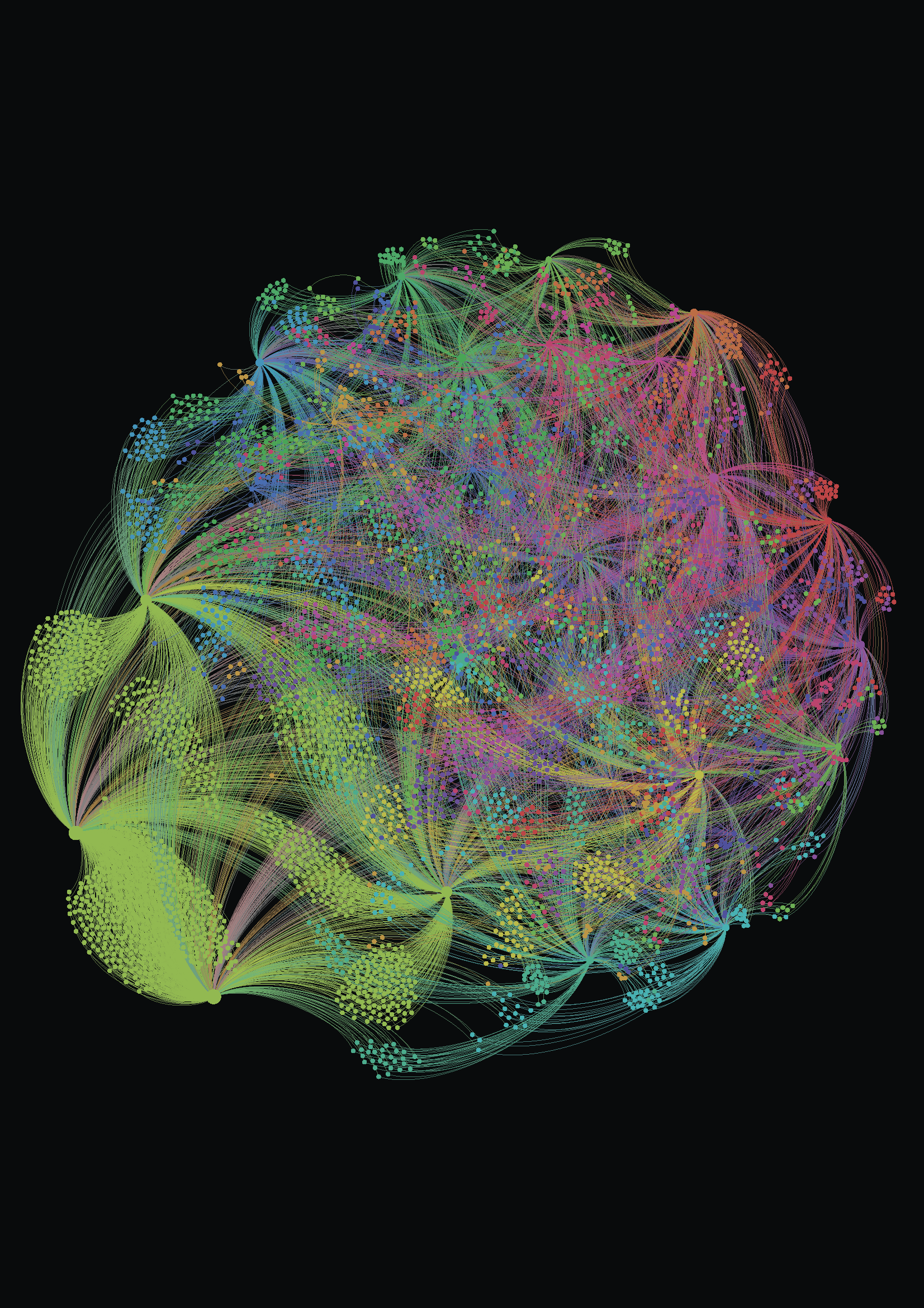}
\caption{A network with 2000 nodes and $r=2\cdot10^5$}
\label{fig:pic-r}
\end{figure}

In real networks with appearing trends, one would expect a dense connected region (group) attached to
each hub, whereas the trending hubs themselves are not tightly connected.
This is another feature which is not observed in PA model. The growing process does not allow for separated groups or clusters, reflecting fans of a particular trend.
However, the TPA model has the ability to generate such patterns. Figure~\ref{fig:pic-r} reveals the structure of a network with a very high tendency to trends.

Obviously, while some networks have high trendiness, others may have almost none ($r \rightarrow 0$).
Let us examine, for example, the word co-occurrence network. The network is formed by the frequency of co-occurrences of words in sentences.  If we consider this network in  the pre-web era, it is clearly a very slow evolving network, as languages evolved very slowly\footnote{It is very possible that with the advance of the web, Twitter and blogs, and the global and inter-cultural interactions  that arrive with them, this assumption is no longer correct. There is evidence that  words's popularity changes fast now, and new words appear quite often. Hence the word co-occurrence network may has just changed its tendency for trends.}. New words appear seldom,  and words' popularity is almost stable over periods of decades.
Thus, a book written over a period of a few years is a static word network, namely with $r=0$, and its word degree distribution is close to a power law.
Bi \etal made an interesting observation \cite{Bi:2001kx} that unlike many other books, the Bible word distribution deviates from a pure Zipf distribution.  Indeed, the Bible, which was written over a few centuries, captures the language dynamics that is responsible for the deviation.

\begin{figure*}
\centering
\begin{tabular}{cc}
\hspace{-.35in}\subfloat[Edits per week ]{\includegraphics[width=7cm]{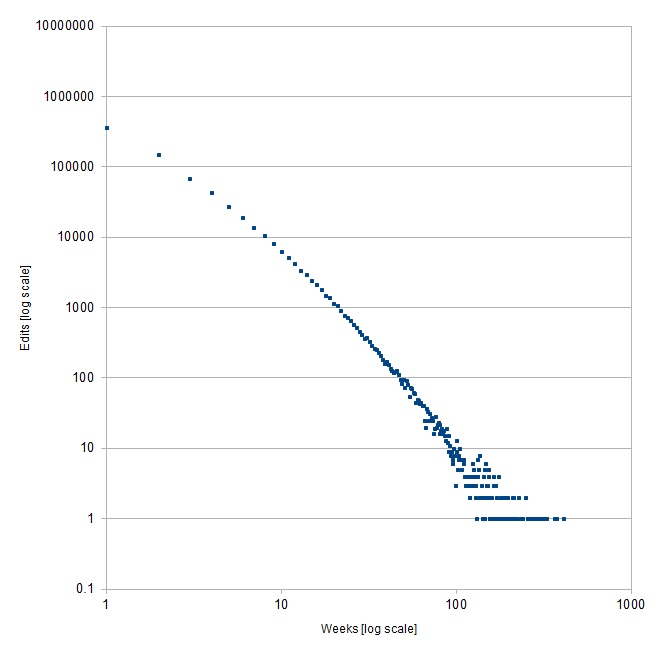}} &
\hspace{-.55in}\subfloat[Filtering trending pages]{\includegraphics[width=6.4cm]{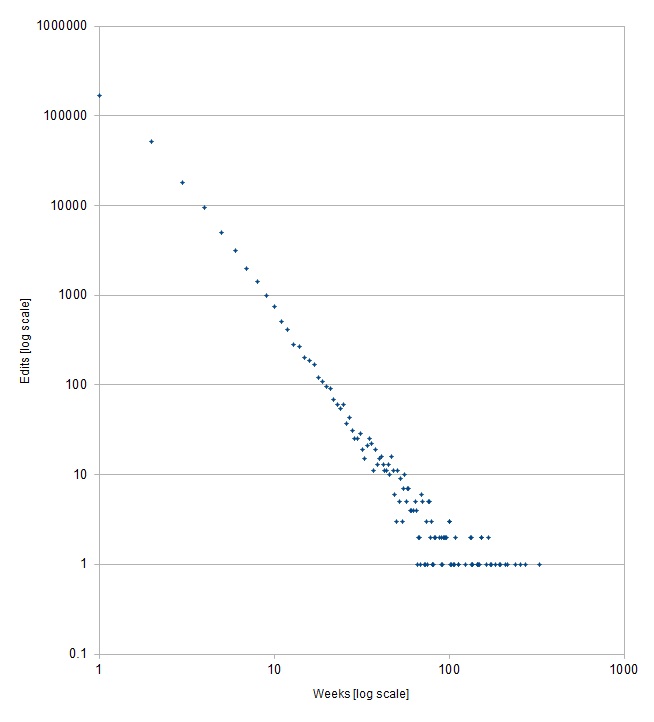}}
\end{tabular}
\caption{Wiki edits per week distribution for  pages created at Feb 25th 2002: (a) features the full distribution of all 8752 pages; (b) features the static part containing the non-trending 6024 pages distribution, where the trending pages are filtered out.}
\label{fig:wikitrends}
\end{figure*}

Thus, we can see the degree distribution of a network as a superposition of its static behavior, governed by a power-law, and the trending dynamics.
To demonstrate this, we investigated the trends in weekly Wiki edit pages  created at a single day. The small deviation of the distribution from a power law suggests a low trendiness.  We filtered out pages in which there were edits in the last 4 weeks. The results, denoted in Figure~\ref{fig:wikitrends} show that without the trending pages the edit distribution is closer to a power law.


%\bibliographystyle{IEEEtran}
%\bibliography{/Users/admin/Dropbox/Research/AllCites}

\begin{thebibliography}{10}
\providecommand{\url}[1]{#1}
\csname url@samestyle\endcsname
\providecommand{\newblock}{\relax}
\providecommand{\bibinfo}[2]{#2}
\providecommand{\BIBentrySTDinterwordspacing}{\spaceskip=0pt\relax}
\providecommand{\BIBentryALTinterwordstretchfactor}{4}
\providecommand{\BIBentryALTinterwordspacing}{\spaceskip=\fontdimen2\font plus
\BIBentryALTinterwordstretchfactor\fontdimen3\font minus
  \fontdimen4\font\relax}
\providecommand{\BIBforeignlanguage}[2]{{%
\expandafter\ifx\csname l@#1\endcsname\relax
\typeout{** WARNING: IEEEtran.bst: No hyphenation pattern has been}%
\typeout{** loaded for the language `#1'. Using the pattern for}%
\typeout{** the default language instead.}%
\else
\language=\csname l@#1\endcsname
\fi
#2}}
\providecommand{\BIBdecl}{\relax}
\BIBdecl

\bibitem{watts1998small}
D.~Watts and S.~Strogatz, ``Collective dynamics of small-world networks,''
  \emph{Nature}, vol. 393, pp. 440--442, 1998.

\bibitem{albert2002statistical}
R.~Albert and A.~Barab{\'a}si, ``Statistical mechanics of complex networks,''
  \emph{Reviews of modern physics}, vol.~74, no.~1, p.~47, 2002.

\bibitem{newman2011structure}
M.~Newman, A.~Barabasi, and D.~Watts, \emph{The structure and dynamics of
  networks}.\hskip 1em plus 0.5em minus 0.4em\relax Princeton University Press,
  2011.

\bibitem{barabasi1999emergence}
A.~Barab{\'a}si and R.~Albert, ``Emergence of scaling in random networks,''
  \emph{science}, vol. 286, no. 5439, pp. 509--512, 1999.

\bibitem{dorogovtsev2000structure}
S.~Dorogovtsev, J.~Mendes, and A.~Samukhin, ``Structure of growing networks
  with preferential linking,'' \emph{Physical Review Letters}, vol.~85, no.~21,
  pp. 4633--4636, 2000.

\bibitem{denison1985trends}
E.~Denison, \emph{Trends in American economic growth, 1929-1982}.\hskip 1em
  plus 0.5em minus 0.4em\relax Brookings Institution Press, 1985.

\bibitem{ng2003major}
F.~Ng and A.~Yeats, ``Major trade trends in east asia: what are their
  implications for regional cooperation and growth?'' \emph{World Bank Policy
  Research Working Paper}, no. 3084, 2003.

\bibitem{mathioudakis2010twittermonitor}
M.~Mathioudakis and N.~Koudas, ``Twittermonitor: trend detection over the
  twitter stream,'' in \emph{Proceedings of the 2010 international conference
  on Management of data}.\hskip 1em plus 0.5em minus 0.4em\relax ACM, 2010, pp.
  1155--1158.

\bibitem{asur2011trends}
S.~Asur, B.~Huberman, G.~Szabo, and C.~Wang, ``Trends in social media:
  Persistence and decay,'' in \emph{5th International AAAI Conference on
  Weblogs and Social Media}, 2011.

\bibitem{Koenigstein:2012fk}
N.~Koenigstein and Y.~Shavitt, ``Talent scouting in p2p networks,''
  \emph{Computer Networks}, vol.~56, no.~3, pp. 970--982, 23 Feb. 2012.

\bibitem{dover2012network}
Y.~Dover, J.~Goldenberg, and D.~Shapira, ``Network traces on penetration:
  Uncovering degree distribution from adoption data,'' \emph{Marketing
  Science}, vol.~31, no.~4, pp. 689--712, 2012.

\bibitem{Bi:2001kx}
Z.~Bi, C.~Faloutsos, and F.~Korn, ``The dgx distribution for mining massive,
  skewed data,'' in \emph{Proceedings of the seventh ACM SIGKDD international
  conference on Knowledge discovery and data mining}.\hskip 1em plus 0.5em
  minus 0.4em\relax ACM, 2001, pp. 17--26.

\end{thebibliography}
\end{document}